Original Research Article

**Title: Prospective Prediction of Future SARS-CoV-2 Infections Using Empirical Data on a National Level to Gauge Response Effectiveness**

Natalia Blanco[1], Kristen Stafford[1,2], Marie-Claude Lavoie[1], Axel Brandenburg[3], Maria W. Górna [4], Matthew Merski[4,†]

[1]Center for International Health, Education, and Biosecurity, Institute of Human Virology - University of Maryland School of Medicine, Baltimore, Maryland USA

[2]Department of Epidemiology and Public Health, University of Maryland School of Medicine, Baltimore, Maryland USA

[3]Nordita, KTH Royal Institute of Technology and Stockholm University, SE-10691 Stockholm, Sweden

[4]Structural Biology Group, Biological and Chemical Research Centre, Department of Chemistry, University of Warsaw, Warsaw, Poland

**Corresponding author(s):** Matthew Merski, Maria Gorna

Address: Biological and Chemical Research Centre, Żwirki i Wigury 101

02-089 Warsaw, Poland

Telephone: (+48) 22 55 26 642

Email: merski@gmail.com; mgorna@chem.uw.edu.pl

**Running head:** Prediction of Future SARS-CoV-2 Infections

**Conflict of interest:** None

**Sources of funding:** The work was supported by the National Science Centre, Poland [grant agreement 2014/15/D/NZ1/00968 to M.W.G.] & the EMBO Installation Grant, European Molecular Biology Organization to M.W.G.

**Data availability:** All data used in this manuscript are publically available at

https://www.ecdc.europa.eu/en/publications-data/download-todays-data-geographic-distribution-covid-19-cases-worldwide

**Acknowledgements:** We thank Dr. Marcin Ziemniak for useful discussions.

**ABSTRACT:**

**Background:** Predicting an accurate expected number of future COVID-19 cases is essential to properly evaluate the effectiveness of any treatment or preventive measure. This study aimed to identify the most appropriate mathematical model to prospectively predict the expected number of cases without any intervention.

**Methods:** The total number of cases for the COVID-19 epidemic in 28 countries was analyzed and fitted to several simple rate models including the logistic, Gompertz, quadratic, simple square, and simple exponential growth models. The resulting model parameters were used to extrapolate predictions for more recent data.

**Results:** While the Gompertz growth models (mean $R^2 = 0.998$) best fitted the current data, uncertainties in the eventual case limit made future predictions with logistic models prone to errors. Of the other models, the quadratic rate model (mean $R^2 = 0.992$) fitted the current data best for 25 (89 %) countries as determined by $R^2$ values. The simple square and quadratic models accurately predicted the number of future total cases 37 and 36 days in advance respectively, compared to only 15 days for the simple exponential model. The simple exponential model significantly overpredicted the total number of future cases while the quadratic and simple square models did not.

**Conclusions:** These results demonstrated that accurate future predictions of the case load in a given country can be made significantly in advance without the need for complicated models of population behavior and generate a reliable assessment of the efficacy of current prescriptive measures against disease spread.

**INTRODUCTION:**

On March 11, 2020 the World Health Organization (WHO) declared the novel coronavirus outbreak (SARS-CoV-2 causing COVID-19) as a pandemic[1] more than three months after the first cases of pneumonia were reported in Wuhan, China in December, 2019[1]. From Wuhan the virus rapidly spread globally, currently leading to ten million confirmed cases and half a million deaths around the world. Although coronaviruses have a wide range of hosts and cause disease in many animals, SARS-CoV-2 is the seventh named member of the *Coronaviridae* known to infect humans[2]. An infected individual will start presenting symptoms an average of 5 days after exposure[3] but approximately 42% of infected individuals remain asymptomatic[4,5]. Furthermore, almost six out of 100 infected patients die globally due to COVID-19[6].

Currently, treatment and vaccine options for COVID-19 are limited[7]. There is currently no effective or approved vaccine for SARS-CoV-2 although a report from April 2020 noted 78 active vaccine projects, most of them at exploratory or pre-clinical stages[8]. As the virus is transmitted mainly from person to person, prevention measures include social distancing, self-isolation, hand washing, and use of masks. Strict measures of quarantine have been shown as the most effective mitigation measures, reducing up to 78% of expected cases compared to no intervention[9]. Nevertheless, to evaluate the actual effectiveness of any mitigation measure it is necessary to accurately predict the expected number of cases in the absence of intervention.

While there has been some early concern about the ability of SARS-CoV-2 to spread at an apparent near exponential rate[10], real limitations in available resources (*i.e.* susceptible population) will reduce the spread to a logistic growth rate[11]. Logistic growth produces a sigmoidal curve (Figure 1) where the total number of cases (**N**) eventually asymptotically approaches the population carrying capacity ($N_M$), which for viral epidemics is analogous to the fraction of the population that will be infected before "herd immunity" is achieved[12,13]. This is represented in derivative form by the generalized logistic function (**Equation 1**):

**Eqn. 1)** $$\frac{dN}{dt} = rN^\alpha \left[1 - \left(\frac{N}{N_M}\right)^\beta\right]^\gamma$$

where **α**, **β**, & **γ** are mathematical shape parameters that define the shape of the curve, and **r** is the general rate term, analogous to the standard epidemiological parameter, $R_0$, the reproductive number, which is a measure of the infectivity of the virus itself [13,14]. For a logistic curve where **α** = ½ and **β** = **γ** = 0, one gets quadratic growth[15] with $N = (rt/2)^2$, while for **α** = **β** = **γ** = 1, this equation can be re-arranged to quadratic form (**Equation 2**)[11] and integrated (**Equation 3**):

**Eqn. 2)** $$\frac{dN}{dt} = \frac{-rN^2}{N_M} + rN$$

**Eqn. 3)** $$N(t) = \frac{N_0 N_M}{(N_M - N_0)e^{-rt} + N_0}$$

where $N_0$ represents the initial number of cases within the population. The shape parameters of logistic functions magnify the uncertainty in fitting these curves especially during the early part of the epidemic. These difficulties are further exacerbated by the additional uncertainty in estimation of $R_0$ for SARS-CoV-2, which is directly linked to the current uncertainty in the population carrying capacity $N_M$. And while the basic logistic function gives rise to a symmetrical sigmoidal curve, asymmetrical curves such as the Gompertz growth function (**Equation 4**)[16,17]:

**Eqn. 4)** $$N(t) = N_0 e^{\left(\ln(\frac{N_M}{N_0})(1 - e^{(-rt)})\right)}$$

which emerges[11] from **Equation 1** for **α** = **γ** = 1, where **r** is replaced by **βr** and **β** approaches 0. With this function the rate of spread slows significantly after passing the mid-point resulting in long-tailed epidemics (Figure 1).

Traditionally the number of cases that will occur in an epidemic like COVID-19 is modeled with an SEIR model (**S**usceptible, **E**xposed, **I**nfected, **R**ecovered/**R**emoved), in which the total population is divided into four categories: susceptible - those who can be infected, exposed – those who in the incubation period but not yet able to transmit the virus to others, infectious - those who are capable of spreading disease to the susceptible population, and recovered/removed – those who have finished the disease course and are not susceptible to re-infection or have died. For a typical epidemic, the ability for infectious individuals to spread the disease is proportional to the fraction of the population in the susceptible category with "herd immunity"[12,13] and extinction of the epidemic occurs once a limiting fraction of the population has entered into the Recovered/Removed category[13]. However, barriers to transmission, either natural[18] or artificial (*i.e.* quarantines, vaccines)[13] can extinguish the epidemic before the community is fully infected. Artificial barriers such as mandatory quarantining in China[19,20] or aggressive contact tracing in South Korea[21] currently seem to have largely stemmed the spread of SARS-CoV-2. Numerous political, social, and material factors prevent the implementation of either of these responses in many other countries[22], but it should be possible to find alternative approaches which can be equally effective. For an epidemic as serious as COVID-19, it behooves medical, scientific, and policy experts to determine as rapidly as possible which community responses are effective and achievable within different populations. However, gauging the effectiveness of these responses requires an accurate prediction[23] of the number of future cases, and overestimation of the expected number of cases will make neutral or even harmful responses appear to be effective when those overestimated cases fail to occur. Thus, accurate prospective predictions (*i.e.* before knowing the actual outcome) are preferable to retrospective analysis in which effectiveness is gauged after the results of the prescriptive actions are known[24,25].

This study aimed to evaluate if a simple model was able to correctly prospectively predict the total number of cases at a future date. We found that fitting the case data to a quadratic (parabolic) rate curve[15] for the early points in the epidemic curves (before the mitigation efforts began to have effects) was easy, efficient, and made good predictions for the number of cases at future dates despite significant national variation in the start of the infection, mitigation response, or economic condition.

**METHODS:** Data on the number of COVID-19 cases was downloaded from the European Centre for Disease Prevention and Control (ECDC) on June 1, 2020[26]. Countries that had reported the highest numbers of cases in mid-March 2020 (and Russia) were chosen as the focus of our analysis to minimize statistical error due to small numbers. The total number of cases for each country was calculated as a simple sum of that day plus all previous days. Days that were missing from the record were assigned values of zero. The early part of the curve was fit and statistical parameters were generated using Prism 8 (GraphPad) using the non-linear regression module using the program standard centered second order polynomial (quadratic), exponential growth, and the Gompertz growth model as defined by Prism 8, and a simple user-defined simple square model ($N = At^2 + C$) where $N$ is the total number of cases, $A$ and $C$ are the fitting constants, and t is the number of days from the beginning of the epidemic curve. The beginning of the curve (SI Table 1) was defined empirically among the first days in which the number of cases began to increase regularly. Typically, this occurred when the country had reported less than 100 total cases. The early part of the curve was defined by manual examination looking for changes in the curve shape and later confirmed by $R^2$ values for the quadratic model. Prospective predictions for the number of cases were done by fitting the total number of COVID-19 cases for each day starting with day 5 and then extrapolating the number of cases using the estimated model parameters to predict the number of cases for the final day for which data was available (June 1, 2020) or to the last day before significant decrease in the $R^2$ value for the quadratic fit. Fit parameters for the Gompertz growth model were not used to make predictions if the fit itself was ambiguous. Acceptable predictions were defined as being within a factor of two from the actual number (*i.e.* predictions within 50-200% of the actual total).

## RESULTS:

**A simple exponential growth model is a poor fit for the SARS-COV-2 pandemic:**

The total number of cases for each of 28 countries was plotted with time and several model equations were fit to the early part of the data before mitigating effects from public health policies began to change the rate of disease spread. In total, 20 (71 %) countries showed mitigation of disease spread by June 1 (Figure 2). When the early, pre-mitigation portion of the data was examined for all 28 countries, the Gompertz growth model had the best statistical parameters (mean $R^2$ = 0.998 ± 0.0028, Table 1) although a fit could not be obtained for the data from 2 countries and many of the fit values for $N_M$ were unrealistic compared to national populations (*e.g.* China and India had predicted $N_M$ values corresponding to 0.014 % and 0.33 % of their populations respectively[26] (SI Table 2)). Fitting was also incomplete for the generalized logistic model for all 28 countries underlining the difficulty in applying this model. On the other hand, the simple models were able to robustly fit all the current data, with the quadratic (parabolic) model performing the best (mean $R^2$ = 0.992 ± 0.004) and the exponential model the worst (mean $R^2$ = 0.957 ± 0.022)(Table 1). In only three (11 %) countries did the exponential model have the best overall $R^2$ value among the simple models. Furthermore, the trend of the overall superiority of the Gompertz model followed by the quadratic was also observed in the standard error of the estimate statistic as well. The mean standard error of the estimate (Sy.x, analogous to the root mean squared error for fits of multiple parameters) value for the 28 countries was 1699 for the Gompertz model, 5613 for the quadratic model, 8572 for the simple square model and 11257 for the exponential model (Table 1). Likewise, plots of the natural log of the total number of cases in the early parts of the epidemic (ln$N$) with time are significantly less linear (as determined by $R^2$) than equivalent plots of the square root of the total number of cases ($N^{1/2}$) (SI Table 3, SI Figs 1, 2).

**Quadratic growth models provide improved fits to the early portion of the epidemic courses:**

While logistic growth models have been widely used to model epidemics[16,27], uncertainties in estimates of **$R_0$** (and therefore the population carrying capacity $N_M$) make prospective predictions of the course of the epidemic difficult[14,27]. (Figure 3, Table 2, SI Table 4). Here we define predictions as accurate when they are within a factor of two (50-200%) of the actual outcome. For most countries, the simple exponential model massively overpredicts the number of future cases. Predictions generated more than 14 days prior were more than double the actual number of cases for 17 (61 %) countries examined. In fact, for 15 (54 %) countries, the exponential model made at least one overprediction by a factor of greater than 10,000 fold, while the quadratic and simple square models make no overprediction by more than a factor of 3.3 and 2.1, respectively (*i.e.* using the first 10 days of data from Portugal the exponential model predicts 34 million cases while the quadratic, simple square, and Gompertz growth models predict 24957, 20358 and 18953 cases respectively while 23683 total cases were observed while the total population of Portugal in 2018 was 10.3 million[26]).

Predictions using the quadratic and simple square models were much more accurate. Only in four (14 %) countries does the quadratic model ever overpredict the final number of cases by more than a factor of two while the simple square model overpredict by a factor of two for only one (4 %) country (SI Table 4). For the quadratic model, the mean maximum daily overprediction was a factor of 1.6-fold (median 1.3 fold) while for the simple square model the mean maximum daily overprediction was 1.3-fold (median 1.1 fold). Both of these models produced much more accurate predictions than the simple exponential model (Table 2).

**DISCUSSION:**

The start of the global SARS-CoV-2 pandemic has resulted in an unprecedented set of national responses. These responses have varied considerably from a strict lockdown in China[10], to aggressive contact tracing in South Korea[28], to mandatory shelter in place restrictions in France[29], to giving citizens information and allowing them more freedom to make choices as in Sweden[30], and to other countries attempting to accelerate their progress towards herd immunity[31]. The variety in these national prescriptions is a result of the different socioeconomic situations in individual countries which have to take into account not only the costs of these efforts both monetarily and in terms of lives, but also what can be reasonably achieved depending on the relationship between individual governments and their citizenry. Additionally, the spread of SARS-CoV-2 has been putatively linked to several inherent factors within a country, such as average population density[32], normal social behaviors[22], and even weather may have an effect[33]. The efficacy of similar prescriptions can vary in pairs of neighboring countries (*i.e.* the UK or Ireland). Therefore, it behooves every nation to review the results of its own policy prescriptions in order to make necessary course adjustments as quickly and accurately as possible.

While predictions about the future course of an epidemic, especially one as novel as COVID-19 are difficult under the best of circumstances, the severity of the pandemic has resulted in an unprecedented amount of epidemiological data being produced with daily frequency. Because logistic epidemic models have been in use for more than a century[16] it is almost tautological that the rate of the spread of SARS-CoV-2 cannot be well-modeled by simple exponential growth equations, further demonstrated by statistical analysis of and the poor predictions made by the exponential model for the future number of cases as compared to the quadratic (parabolic) model[15]. However, simple exponential models do not generate entirely terrible fit statistics (Table 1), and this may account for the conflation of the course of the SARS-CoV-2 pandemic with truly exponential growth. That the exponential growth constant term, **k**, is constantly decreasing after day 10 in 10 (68 %) countries (SI Fig. 3) further indicates the overall utility of logistic models, which were explicitly developed to model the a constantly decreasing rate of growth due to consumption of the available resource (*i.e.* the susceptible population pool of the SIR model)[16]. But, while logistic models are implicitly the correct model, they are difficult to accurately fit during the early portion of an epidemic due to inherent uncertainties in the mathematical shape parameters (Equation 1) of the curve itself and the population carrying capacity for SARS-CoV-2, $N_M$, which still has a significant uncertainty as the virus has only recently moved into the human population. Herd immunity is defined as $1 - 1/\mathbf{R_0}$, and since current estimates for $\mathbf{R_0}$ vary from $1.5 - 6.5$[14]. This implies that $33 - 85\%$ of the population will need to have contracted the disease and developed immunity in order to terminate the epidemic. A discrepancy of this size will significantly affect predictions based on logistic growth models.

Here we note the utility of the quadratic (parabolic) and simple square models in predicting the course of the pandemic more than a month in advance. The simple exponential model vastly overpredicts the number of cases (Fig. 2, Table 2). The Gompertz growth model, while often making largely correct predictions often generates wildly inaccurate estimates of the population carrying capacity $N_M$ (SI Table 2), and the generalized logistic model simply fails to produce a statistically reliable result with the currently available data. Overestimation of the future number of cases will cause problems because the failure of the number of predicted cases to materialize may be erroneously used as evidence that poorly implemented and ineffective policy prescriptions are reducing the spread of SARS-CoV-2, which may lead to political pressure for premature cessation of all prescriptive measures and inevitably an increase in the number of cases and excess, unnecessary morbidities. Fortunately, the quadratic model produces accurate, prospective predictions of the number of cases (Fig. 3, Table 2). Use of this model is simple as it is directly implemented in common spreadsheet programs and can be implemented without much difficulty or technical modeling expertise. In theory, this model can also be applied to smaller, sub-national populations, although the smaller number of total cases in these regions will undoubtedly give rise to larger statistical errors.

In no way does the empirical agreement between the quadratic model and empirical data negate the fact that the growth of the SARS-CoV-2 epidemic is logistic in nature in all 28 countries (Table 1, SI Table 2). We expect the suitability of these empirical quadratic fits is related to either the fact that quadratic form of the slope of the generalized logistic function or the limitation of the virus to a physical radius of infectivity around infectious individuals, or that it is still early in the pandemic as no country has yet officially logged even 1% of its population as having been infected, or all three. Of course, the true number of COVID-19 cases is a matter of debate as there is speculation that a significant fraction of infections are not being identified[34]. However, because this method is focused on the rate of case growth over time, the errors that lead to any undercounting within a given country are likely to remain largely unchanged over the short time periods observed here and still provide a reasonable estimate of the number of positively identified cases. While despite their similar predictive power we largely focus on the quadratic model rather than the simple square model for the aforementioned reasons, we must also note that quadratic curve fitting is natively implemented in most common spreadsheet software while the simple square model is not. By monitoring the $R^2$ values for the quadratic models, it is a simple task to identify when the epidemic is beginning to subside within a country (*i.e.* "bending the curve"). Here we recommend the use of an $R^2$ value of 0.985 for identifying when the rate of infection is beginning to subside, but more conservative estimates can also be made by lowering this threshold.

Examination of the data collected here suggests that early, aggressive measures have been most effective at reducing disease burden within a country. Countries that initially adopted less stringent measures (such as the US, UK, Russia, and Brazil) are currently more heavily burdened than those countries that started with more intense prescriptions (such as China, South Korea, Australia, Denmark, and Vietnam)[35]. Vietnam was not analyzed here as it does not currently have a large number of COVID-19 cases likely due to its early aggressive action against viral spread[7,35]. The effectiveness of aggressive measures may also be due to the apparent quadratic rate of growth of total cases with time (Equation 2); while growth in proportion to the square of the number of days is fast, it is not as fast as exponential growth. Early reductions in the number of infected individuals and the number of interactions they have with susceptible individuals clearly pays compounded dividends in future case reductions as advantage can be taken of this slower spreading rate.

Quadratic modeling of the increases of COVID-19 cases within national boundaries currently gives a more accurate prospective prediction of the growth in the number of total cases. This allows for near real time adjustments to policy prescriptions and obviates the need to estimate the effect of human behaviors on these predictions and instead focus on the available empirical data and fine tune the needed responses and efficiently guide the direction of health care resources. Until an effective vaccine or cure is available, social distancing, contact tracing, and other aggressive quarantine measures are the most effective tools to combat the spread of SARS-CoV-2 and it is imperative to monitor whether these measures are being effectively implemented. Accurate prospective modeling of the future number of cases within countries will help to minimize the social costs and financial burdens of these necessary mitigation measures.

**TABLES:**

**Table 1:** Statistical fit parameters for the quadratic, simple square, simple exponential, and Gompertz growth models for the early course of the epidemic in each country.

|  |  | Mean | Strd. Dev | Median |
|---|---|---|---|---|
| **Quadratic[a]** (N=28) | **R squared** | 0.9917 | 0.0044 | 0.9930 |
|  | **Sum of Squares** | 21916684140 | 1.05689E+11 | 87110317.5 |
|  | **Sy.x** | 5612.9 | 15068.7 | 1250.0 |
| **Simple Square** (N=28) | **R squared** | 0.9780 | 0.0159 | 0.9821 |
|  | **Sum of Squares** | 44700980001 | 2.02519E+11 | 174119853 |
|  | **Sy.x** | 8572.1 | 21235.6 | 1591.5 |
| **Simple Exponential** (N=28) | **R squared** | 0.9568 | 0.0215 | 0.9548 |
|  | **Sum of Squares** | 1.0317E+11 | 5.23176E+11 | 466907869.5 |
|  | **Sy.x** | 11257.0 | 32734.8 | 3067.0 |
| **Gompertz growth** (N=26) | **R squared** | 0.9978 | 0.0028 | 0.9991 |
|  | **Sum of Squares** | 2048425776 | 9767369361 | 11600080 |
|  | **Sy.x** | 1699.2 | 4598.3 | 459.7 |

[a]Because fit parameters were unable to be obtained for the Gompertz model for two countries, the statistical parameters were calculated from the remaining 26 countries (indicated by $N = 26$). The generalized logistic model failed to adequately determine fit parameters in every cases so is not included in this table.

**Table 2:** Results of prospective predictions of total case load made using the various models.

| | Mean[b] Acceptable Predictive Length (Days) | Mean Minimum Normalized Prediction (Observed/Actual Total Cases) | Mean Maximum Normalized Prediction (Observed/Actual Total Cases) |
|---|---|---|---|
| **Simple Exponential** | 14.6 ± 8.0 (11) | 1.11 ± 0.09 (1.14) | 8.9 x $10^{14}$ ± 4.7 x $10^{15}$ (5 x $10^4$) |
| **Quadratic (Parabolic)** | 35.6 ± 15.8 (31.5) | 0.25 ± 0.29 (0.09) | 1.55 ± 0.64 (1.33) |
| **Simple Square** | 36.9 ± 21.6 (31) | 0.42 ± 0.34 (0.24) | 1.26 ± 0.35 (1.12) |
| **Gompertz Growth** | 28.2 ± 17.9 (25.5) | 0.14 ± 0.24 (0.02) | 3.8 x $10^8$ ± 2.0 x $10^9$ (2 x $10^2$) |

[b] All data given are mean results ± the sample standard error. Median values are given in parentheses.

**FIGURES:**

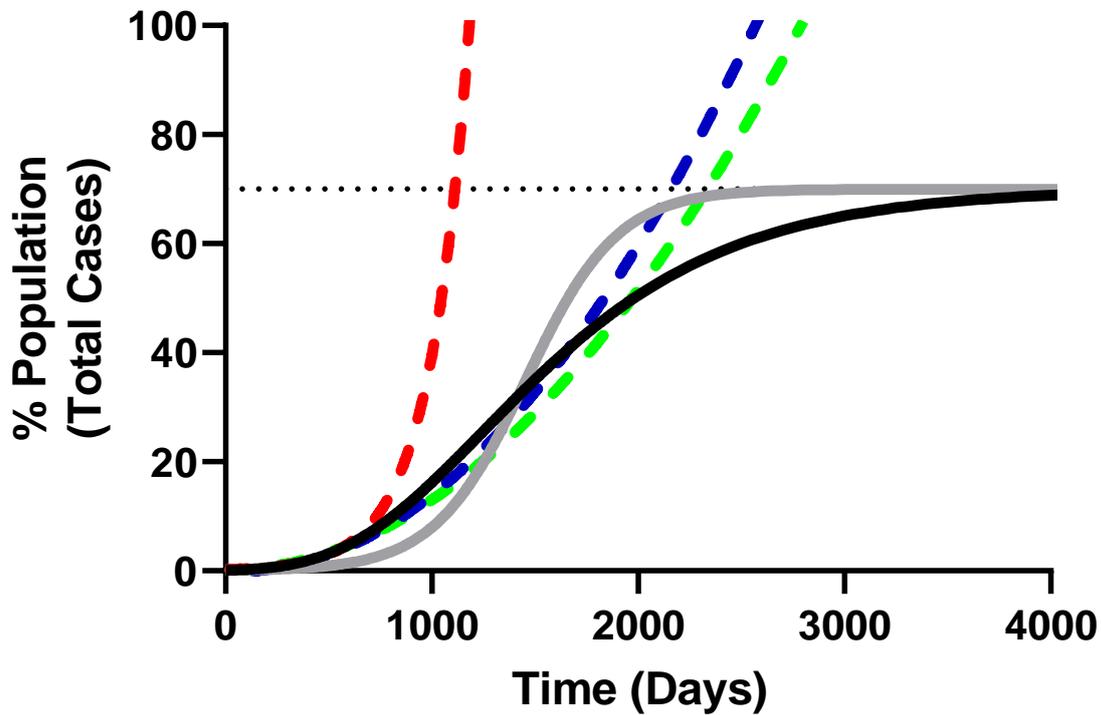

**Figure 1:** Illustrative comparison of exponential, quadratic, generalized logistic and Gompertz growth curves. The Gompertz growth curve (Equation 4, solid black line) representing the progress of a theoretical epidemic for a disease with an arbitrarily chosen **$R_0$** value of 3.4 (**r** = 0.045, **$N_0$** = 1, **$N_M$** = 70%, dotted line). The solid grey line is an equivalent logistic curve, note that while the midpoint for both logistic curves is the same, the Gompertz curve reaches the population carrying capacity more slowly, resulting in a long tailed epidemic. The initial part of the Gompertz curve (including time points until 5% of the population has been infected) was fit to the simple exponential (red dashes), quadratic (blue dashes) and simple square (green dashes) models. It is apparent from these curves how quickly the exponential curve overestimates the rate of growth for the epidemic as compared to the quadratic and simple square fit curves and how the quadratic model more closely follows the Gompertz growth curve, evidenced by the smaller Sy.x value for the quadratic fit in Table 1.

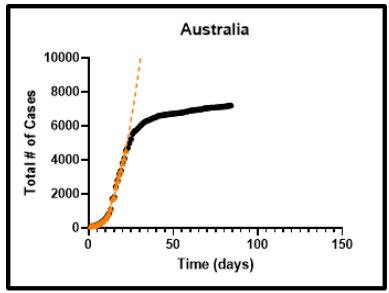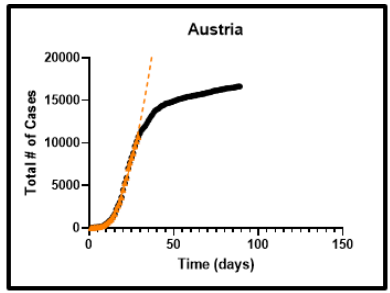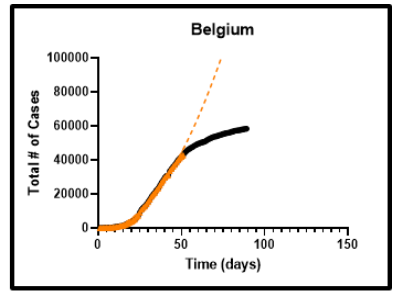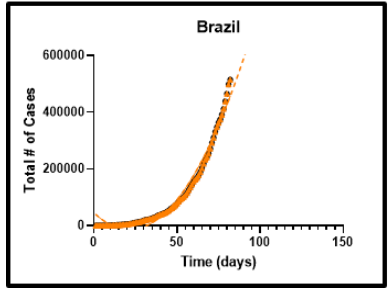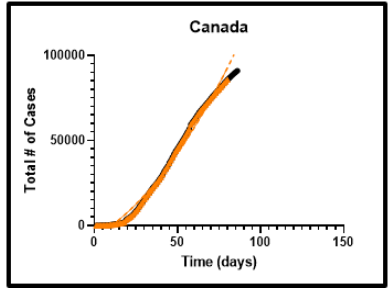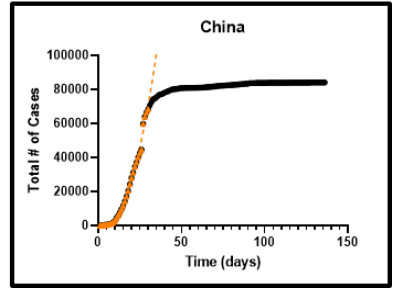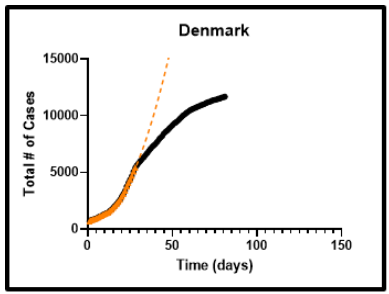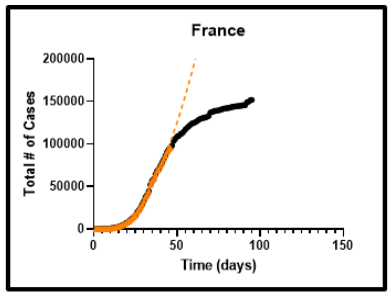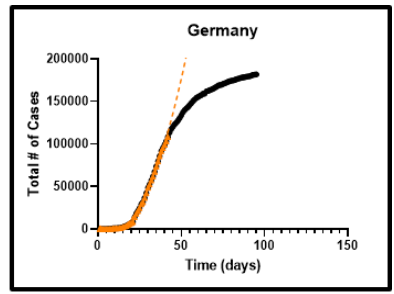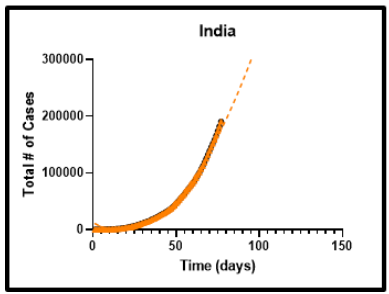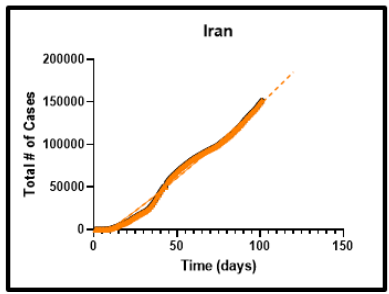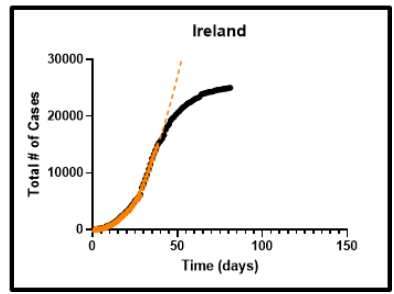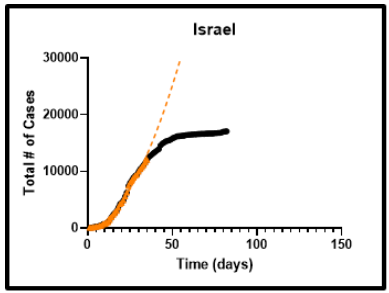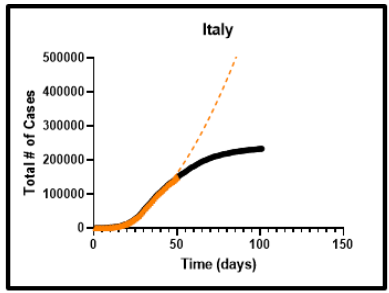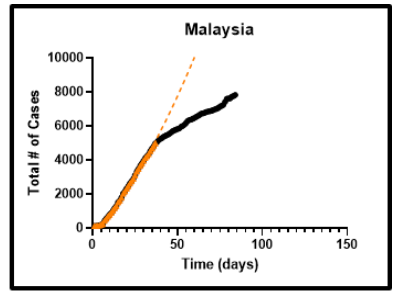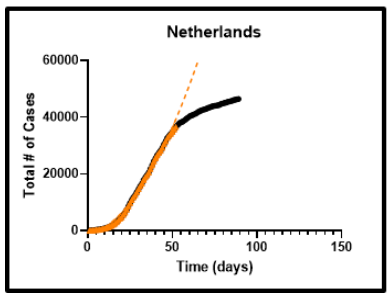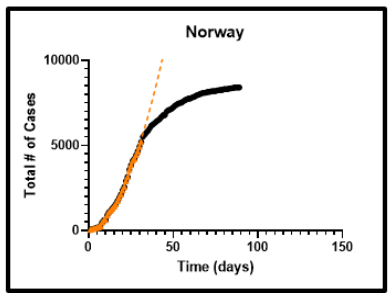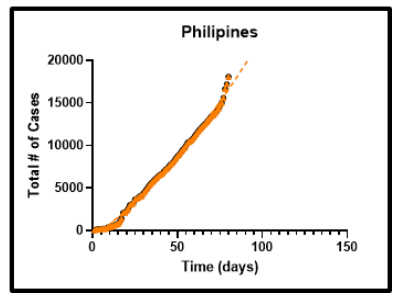

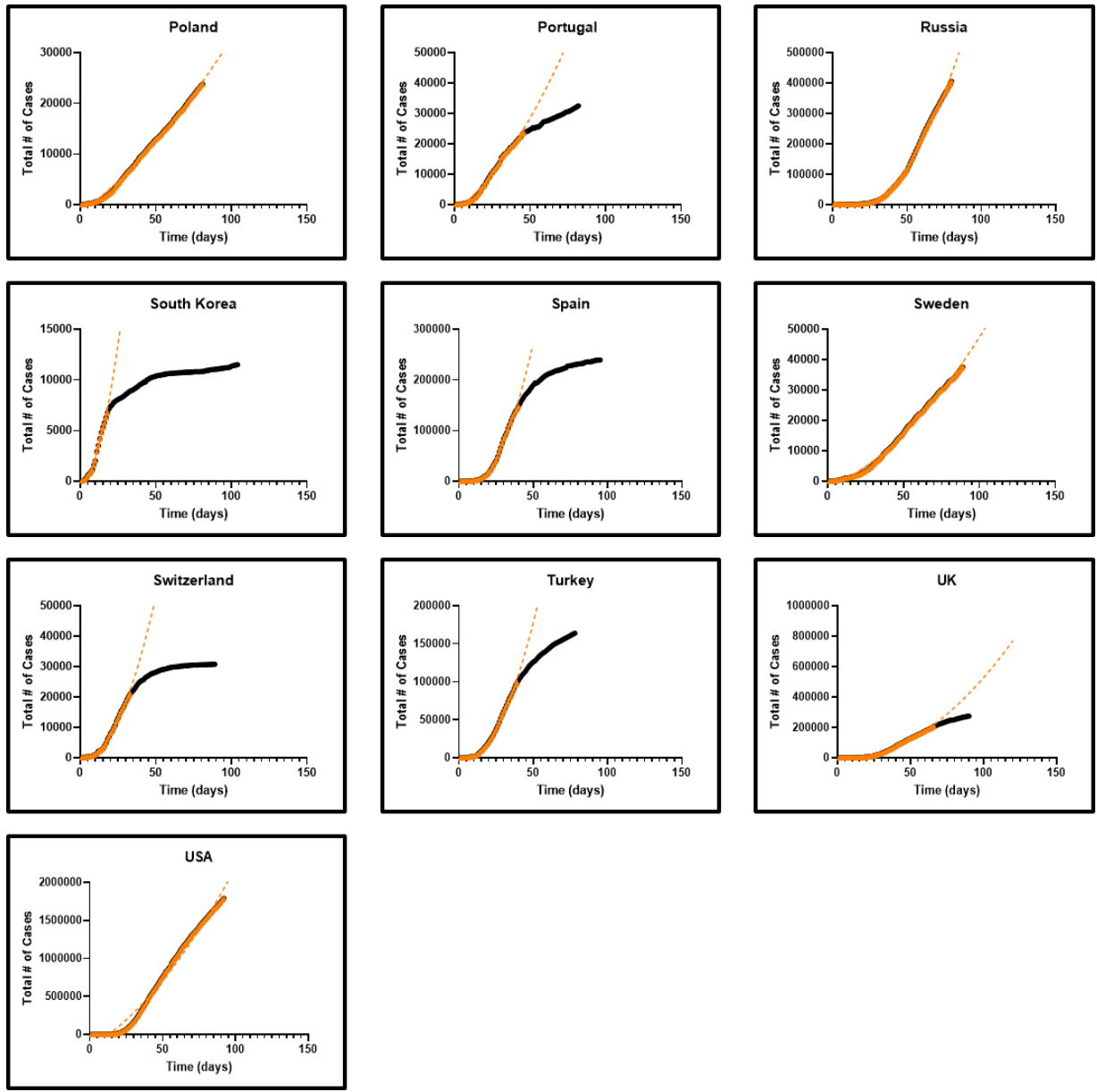

**Figure 2:** The development of COVID-19 cases over time in 28 nations. The total number of cases as of June 1, 2020 is indicated by black circles while the early part of the curve is indicated by orange triangles. A quadratic fit curve based on the early part of the curve extrapolated into the future is shown as an orange dashed line. The black circles are obscured in those countries which had not begun to effectively reduce SARS-CoV-2 spread by June 1, 2020.

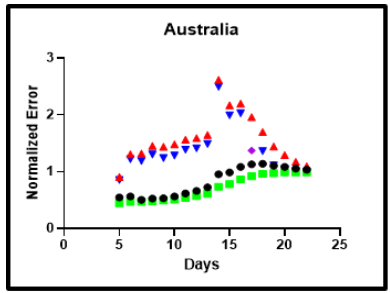
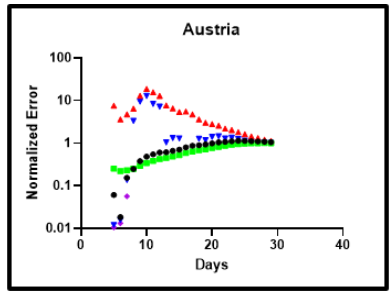
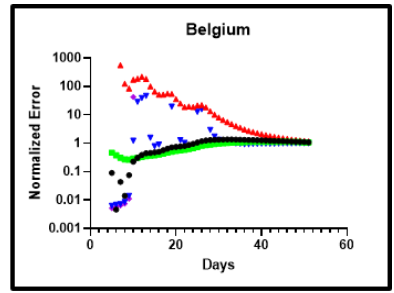
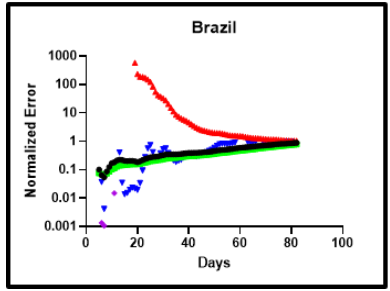
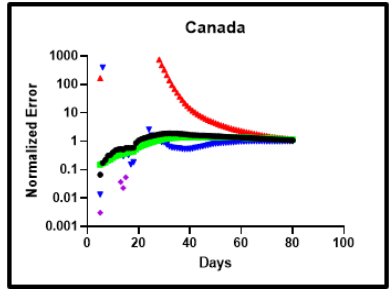
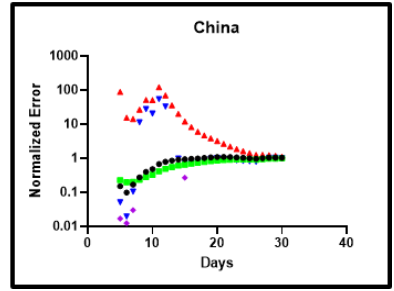
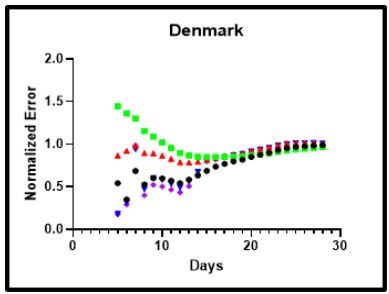
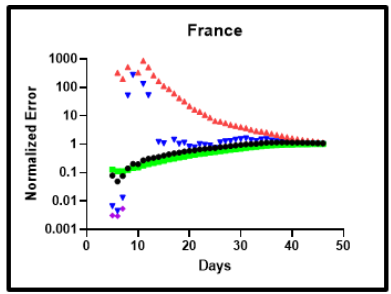
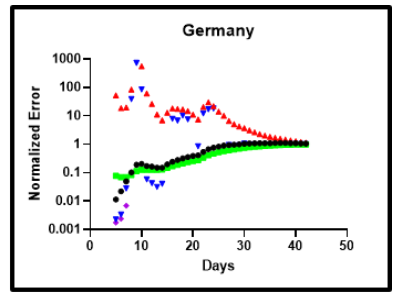
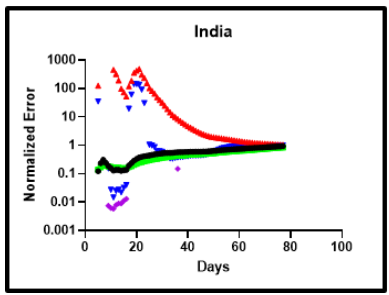
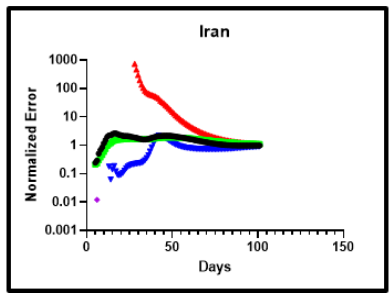
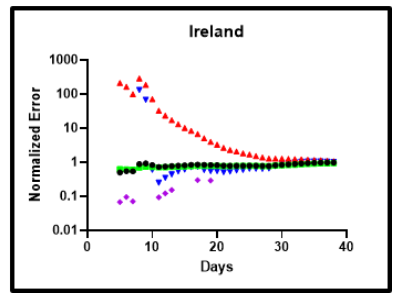
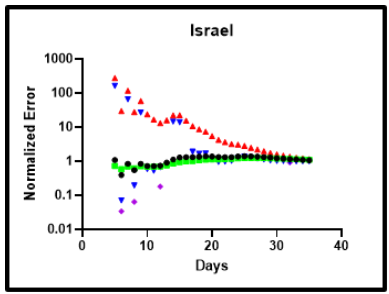
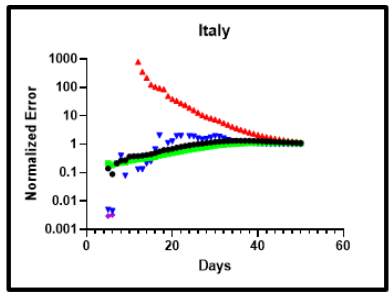
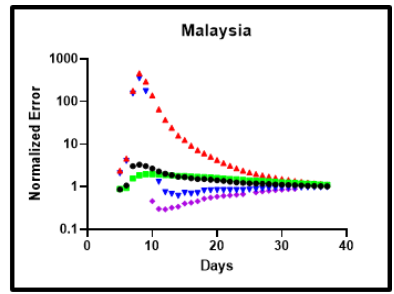
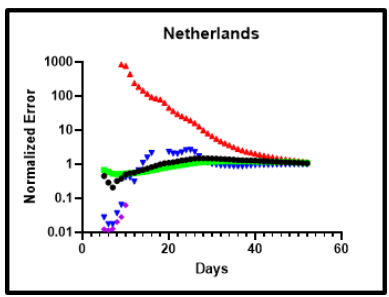
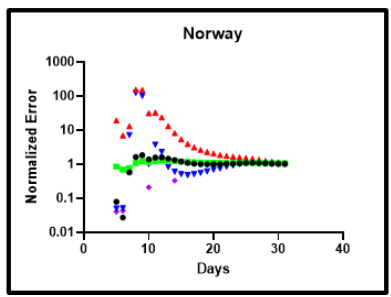
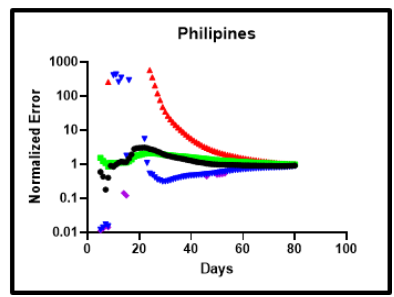

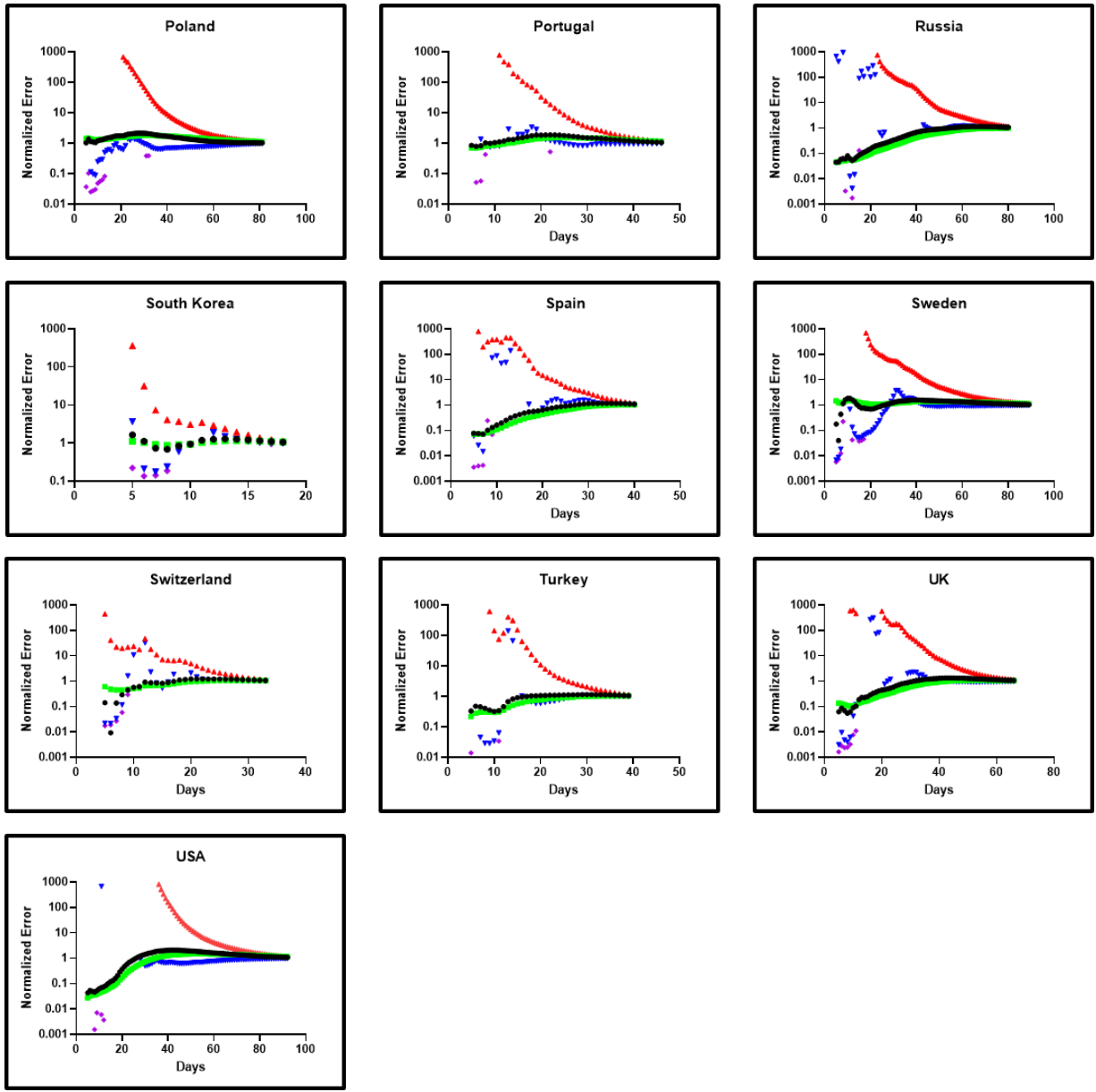

**Figure 3:** Comparison of the errors in prospective predictions for COVID-19 case numbers for different growth models for 28 countries for the simple exponential model (red triangles), the simple square model (green squares), the quadratic model (black circles), the Gompertz growth model (blue triangles), and the basic logistic growth model (purple diamonds). Note the log scale for the vertical axis which indicates the ratio of the predicted to observed number of cases. In each graph the fit values for each model using only data up to that day are used to predict the number of expected cases for the last day for which data is available (or the last day before significant curve deviation is observed, see figure 2). Days on which the fit was not statistically sound were omitted from the graph.

**Supporting Information for "Prospective Prediction of Future SARS-COV-2 Infections Using Empirical Data on a National Level to Gauge Response Effectiveness"** Blanco *et al.*

**SI Table 1:** The start date (day 1, in 2020) of the fit curves for the 28 analyzed countries.

| Country | Date | Country | Date |
|---|---|---|---|
| Australia | Mar. 10 | Italy | Feb. 22 |
| Austria | Mar. 5 | Malaysia | Mar. 10 |
| Belgium | Mar. 5 | Norway | Mar. 5 |
| Brazil | Mar. 12 | Philippines | Mar. 14 |
| Canada | Mar. 8 | Poland | Mar. 13 |
| China | Jan. 18 | Portugal | Mar. 12 |
| Denmark | Mar. 13 | Russia | Mar. 14 |
| France | Feb. 28 | S. Korea | Feb. 19 |
| Germany | Feb. 28 | Spain | Feb. 28 |
| Holland | Mar. 5 | Sweden | Mar. 5 |
| India | Mar. 17 | Swiss | Mar. 5 |
| Iran | Feb. 22 | Turkey | Mar. 16 |
| Ireland | Mar. 13 | UK | Mar. 4 |
| Israel | Mar. 12 | US | Mar. 2 |

**SI Table 2:** The fit parameters for the development of the early portion of the SARS-CoV-2 epidemic in 28 countries for the exponential, quadratic, simple square, simple exponential, Gompertz growth models as calculated for each individual day during the early portion of the epidemic.[a]

| | Exponential | | | | | | Parabolic (Quadratic) | | | | | | |
|---|---|---|---|---|---|---|---|---|---|---|---|---|---|
| | $N_0$ | k | Days | $R^2$ | Sum of Sq. | Sy.x | A | B | C | Days | $R^2$ | Sum of Sq. | Sy.x |
| Australia | 179 | 0.1513 | 20 | 0.9697 | 1425602 | 267 | 13.37 | -90.68 | 239.2 | 19 | 0.9922 | 366492 | 138.9 |
| Austria | 337.8 | 0.1236 | 27 | 0.9659 | 12519859 | 681 | 19.58 | -189.3 | 401.2 | 26 | 0.9943 | 2085220 | 283.2 |
| Belgium | 2142 | 0.06171 | 49 | 0.9511 | 520986359 | 3261 | 19.28 | -57.99 | -864.4 | 48 | 0.9917 | 88470385 | 1358 |
| Brazil | 4256 | 0.05894 | 80 | 0.9969 | 4958718241 | 7873 | 127.9 | -5475 | 45150 | 79 | 0.9783 | 34416594185 | 20872 |
| Canada | 7125 | 0.0332 | 78 | 0.931 | 4749609383 | 7803 | 8.584 | 554.8 | -6616 | 77 | 0.9881 | 817501344 | 3258 |
| China | 2268 | 0.1174 | 28 | 0.973 | 396258176 | 3762 | 106.7 | -897 | 1697 | 27 | 0.9934 | 97656280 | 1902 |
| Denmark | 590.7 | 0.07944 | 26 | 0.9979 | 115592 | 66.68 | 7.066 | -39.85 | 915.5 | 25 | 0.9966 | 186104 | 86.28 |
| France | 2803 | 0.07993 | 44 | 0.9631 | 1712988947 | 6240 | 71.19 | -1116 | 3219 | 43 | 0.9939 | 281922809 | 2561 |
| Germany | 2493 | 0.09302 | 40 | 0.9644 | 1890621844 | 6875 | 104 | -1773 | 5387 | 39 | 0.9947 | 279852332 | 2679 |
| India | 2864 | 0.05511 | 75 | 0.9949 | 1111015939 | 3849 | 49.77 | -1712 | 12620 | 74 | 0.99 | 2195383041 | 5447 |
| Iran | 17024 | 0.02264 | 99 | 0.9188 | 18388322453 | 13629 | 2.445 | 1363 | -13506 | 98 | 0.9874 | 2848350547 | 5391 |
| Ireland | 503.7 | 0.09082 | 36 | 0.9912 | 6518863 | 425.5 | 13.55 | -149.8 | 686 | 35 | 0.9942 | 4253336 | 348.6 |
| Israel | 852.5 | 0.07983 | 33 | 0.9384 | 36190773 | 1047 | 8.56 | 87.1 | -555.6 | 32 | 0.9853 | 8641844 | 519.7 |
| Italy | 6869 | 0.0645 | 48 | 0.946 | 6870075125 | 11964 | 74.09 | -438.2 | -2215 | 47 | 0.9887 | 1434602095 | 5525 |
| Malysia | 594.7 | 0.06073 | 35 | 0.9434 | 5207926 | 385.7 | 1.131 | 104.1 | -268 | 34 | 0.9959 | 376542 | 105.2 |
| Netherlands | 2433 | 0.05498 | 50 | 0.9473 | 412829380 | 2873 | 12.08 | 158 | -1648 | 49 | 0.9918 | 63939110 | 1142 |
| Norway | 375.5 | 0.08815 | 29 | 0.9685 | 2642459 | 301.9 | 4.622 | 31.92 | -62.81 | 28 | 0.9965 | 289515 | 101.7 |
| Philipines | 1773 | 0.02925 | 78 | 0.951 | 100592449 | 1136 | 0.9324 | 142.5 | -775 | 77 | 0.9949 | 10527292 | 369.8 |
| Poland | 2510 | 0.0293 | 79 | 0.9416 | 269292285 | 1846 | 1.284 | 215.8 | -1582 | 78 | 0.9959 | 18900782 | 492.3 |
| Portugal | 2266 | 0.05429 | 44 | 0.9239 | 225040953 | 2262 | 5.106 | 356.9 | -2129 | 43 | 0.9865 | 39799359 | 962.1 |
| Russia | 10951 | 0.04698 | 78 | 0.9621 | 53229872738 | 26123 | 100.9 | -2839 | 13209 | 77 | 0.9947 | 7389136781 | 9796 |
| South Korea | 446.3 | 0.1577 | 16 | 0.9513 | 4604176 | 536.4 | 20.11 | 47.27 | -179.3 | 15 | 0.9892 | 1025210 | 261.4 |
| Spain | 4200 | 0.09283 | 38 | 0.9583 | 4155042694 | 10457 | 149.9 | -2126 | 5154 | 37 | 0.9926 | 737759799 | 4465 |
| Sweden | 3227 | 0.02907 | 87 | 0.9453 | 717453607 | 2872 | 2.874 | 207.7 | -2081 | 86 | 0.9935 | 85750250 | 998.5 |
| Switzerland | 1053 | 0.09459 | 31 | 0.9599 | 64146352 | 1438 | 22.55 | -63.51 | -159.6 | 30 | 0.9943 | 9138733 | 551.9 |
| Turkey | 4164 | 0.08443 | 37 | 0.9649 | 1435404688 | 6229 | 82.09 | -532 | -64.87 | 36 | 0.997 | 123006298 | 1848 |
| UK | 9411 | 0.04888 | 64 | 0.9503 | 15356148331 | 15490 | 56.77 | -360.4 | -3859 | 63 | 0.9917 | 2550481471 | 6363 |
| USA | 149327 | 0.0288 | 90 | 0.9205 | 2.77214E+12 | 175504 | 123.2 | 11347 | -164607 | 89 | 0.9839 | 5.60161E+11 | 79334 |

| | Simple Square | | | | | | Gompertz | | | | | | |
|---|---|---|---|---|---|---|---|---|---|---|---|---|---|
| | A | C | Days | $R^2$ | Sum of Sq. | Sy.x | $N_M$ | $N_0$ | r | Days | $R^2$ | Sum of Sq. | Sy.x |
| Australia | 9.655 | -162.3 | 20 | 0.9834 | 781794 | 197.7 | 9412 | 2.116 | 0.1114 | 19 | 0.9946 | 254929 | 115.8 |
| Austria | 13.64 | -685.4 | 27 | 0.9828 | 6331251 | 484.2 | 19048 | 0.2013 | 0.1041 | 26 | 0.9985 | 556075 | 146.2 |
| Belgium | 18.23 | -1437 | 49 | 0.9915 | 90706366 | 1361 | 60816 | 3.066 | 0.06539 | 48 | 0.9994 | 5988360 | 353.2 |
| Brazil | 65.94 | -40693 | 80 | 0.9252 | 1.18501E+11 | 38487 | N/A | N/A | N/A | N/A | N/A | N/A | N/A |
| Canada | 15.01 | 1875 | 78 | 0.9765 | 1618735064 | 4556 | 107690 | 66.56 | 0.04306 | 77 | 0.9995 | 31367493 | 638.3 |
| China | 79.49 | -3621 | 28 | 0.9862 | 203485104 | 2696 | 196544 | 89.89 | 0.06696 | 27 | 0.9928 | 105567598 | 1977 |
| Denmark | 5.772 | 694.2 | 26 | 0.9935 | 355042 | 116.9 | N/A | N/A | N/A | N/A | N/A | N/A | N/A |
| France | 48.87 | -6743 | 44 | 0.9809 | 886761649 | 4489 | 157199 | 0.2668 | 0.0721 | 43 | 0.9989 | 53127885 | 1112 |
| Germany | 65.21 | -9111 | 40 | 0.9729 | 1436844105 | 5993 | 176314 | 0.02011 | 0.08326 | 39 | 0.9994 | 33044849 | 920.5 |
| India | 29.16 | -12625 | 75 | 0.9588 | 8996060726 | 10952 | 4520256 | 512.2 | 0.01367 | 74 | 0.9997 | 56663182 | 875.1 |
| Iran | 14.99 | 12720 | 99 | 0.9442 | 12629718373 | 11295 | 168414 | 1000 | 0.03245 | 98 | 0.9899 | 2291948974 | 4836 |
| Ireland | 9.934 | -426.6 | 36 | 0.986 | 10338724 | 535.9 | 207497 | 158 | 0.02653 | 35 | 0.9979 | 1584982 | 212.8 |
| Israel | 10.84 | 42.4 | 33 | 0.9826 | 10242445 | 557.1 | 15802 | 3.447 | 0.09749 | 32 | 0.9978 | 1285165 | 200.4 |
| Italy | 66.01 | -6457 | 48 | 0.9878 | 1554850288 | 5691 | 196177 | 0.6338 | 0.0756 | 47 | 0.9996 | 44675513 | 975 |
| Malysia | 3.708 | 485.6 | 35 | 0.9665 | 3085369 | 296.9 | 6837 | 81.38 | 0.06997 | 34 | 0.9991 | 84400 | 49.82 |
| Netherlands | 14.88 | -59.91 | 50 | 0.9896 | 81537845 | 1277 | 51192 | 25.42 | 0.05972 | 49 | 0.9996 | 3208366 | 255.9 |
| Norway | 5.561 | 132.4 | 29 | 0.9948 | 437677 | 122.9 | 11096 | 55.28 | 0.06296 | 28 | 0.9975 | 208669 | 86.33 |
| Philipines | 2.585 | 1406 | 78 | 0.9691 | 63405902 | 901.6 | 25661 | 384.4 | 0.0274 | 77 | 0.9904 | 19675484 | 505.5 |
| Poland | 3.754 | 1761 | 79 | 0.9686 | 144754602 | 1354 | 31792 | 265.5 | 0.03332 | 78 | 0.9968 | 14757805 | 435 |
| Portugal | 12.25 | 1058 | 44 | 0.9656 | 101689588 | 1520 | 28171 | 27.63 | 0.07722 | 43 | 0.999 | 2858438 | 257.8 |
| Russia | 67.96 | -30237 | 78 | 0.9798 | 28369487572 | 19071 | 635427 | 0.3143 | 0.04345 | 77 | 0.9997 | 456703595 | 2435 |
| South Korea | 22.46 | -5.497 | 16 | 0.9885 | 1085625 | 260.5 | 10392 | 6.096 | 0.1595 | 15 | 0.9969 | 293277 | 139.8 |
| Spain | 101.1 | -11435 | 38 | 0.9782 | 2171487122 | 7559 | 223586 | 0.01913 | 0.09264 | 37 | 0.9992 | 75137515 | 1425 |
| Sweden | 5.041 | 1448 | 87 | 0.9816 | 240743441 | 1663 | 51084 | 148.9 | 0.0322 | 86 | 0.9994 | 7670122 | 298.6 |
| Switzerland | 20.79 | -571.9 | 31 | 0.9938 | 9849358 | 563.7 | 33855 | 7.933 | 0.08685 | 30 | 0.9993 | 1091085 | 190.7 |
| Turkey | 69.58 | -4117 | 37 | 0.995 | 206131636 | 2360 | 174393 | 32.93 | 0.06972 | 36 | 0.9998 | 8442355 | 484.3 |
| UK | 51.72 | -8430 | 64 | 0.9911 | 2739402810 | 6542 | 281089 | 5.477 | 0.05277 | 63 | 0.9995 | 156794091 | 1578 |
| USA | 237.8 | 34575 | 90 | 0.9693 | 1.07155E+12 | 109115 | 2069510 | 455.9 | 0.04205 | 89 | 0.9986 | 49886079970 | 23675 |

[a] The fit equations for each are as follows:

Simple exponential: $N = N_0 e^{kt}$

Quadratic: $N = At^2 + Bt + C$

Simple square: $N = At^2 + C$

Gompertz growth: $$N(t) = N_0 e^{\left(\ln(\frac{N_M}{N_0})(1-e^{(-rt)})\right)}$$

where *N* is the total number of cases, t is the time in days, *N₀* is the initial seeding population of the epidemic, $N_M$ is the population carrying capacity (the amount of the population that must be infected to achieve herd immunity), *A*, *B* & *C* are the standard quadratic terms (or for the simple square model equation). Additionally, the number of days of data used in the fitting, the $R^2$, sum of squares, and Sy.x statistical values are given. For the Gompertz growth model, an adequate fit could not be achieved for Brazil or Denmark and this is indicated by *N/A*.

**SI Table 3:** Summary of the statistical parameters of linear fits to the plots of *N* and ln*N* with time.

| Square root of total cases | Mean | Str. Dev. | Median |
|---|---|---|---|
| Slope | 5.639464 | 3.648843 | 4.355 |
| Y-intercept | -21.7129 | 35.10683 | -8.081 |
| X-intercept | 1.442404 | 5.604858 | 1.8695 |
| 1/slope | 0.261796 | 0.166244 | 0.2302 |
| R squared | 0.977639 | 0.012277 | 0.98075 |
| Sy.x | 14.90246 | 17.01295 | 7.7295 |
| **Natural log of total cases** | Mean | Str. Dev. | Median |
| Slope | 0.1353 | 0.057693 | 0.1251 |
| Y-intercept | 5.492857 | 0.82118 | 5.435 |
| X-intercept | -51.5525 | 30.04849 | -42.905 |
| 1/slope | 8.937607 | 4.159834 | 8.014 |
| R squared | 0.891386 | 0.068346 | 0.9139 |
| Sy.x | 0.624629 | 0.278802 | 0.63595 |
| SQRT | Mean | Str. Dev. | Median |

**SI Table 4:** The range of prediction results for the quadratic, simple square, simple exponential, and Gompertz growth models based on days of good predictions before the target, last day of observed data, inclusive.[b]

| | Days of Good predictions | | | | Minimum prediction for final day | | | | Maximum prediction for final day | | | |
|---|---|---|---|---|---|---|---|---|---|---|---|---|
| | Quad | Square | Exp | Gompertz | Quad | Square | Exp | Gompertz | Quad | Square | Exp | Gompertz |
| Australia | 18 | 13 | 6 | 5 | 0.503135 | 0.447172 | 0.906785 | 0.854725 | 1.137994 | 0.989844 | 2.616133 | 2.49387 |
| Austria | 19 | 15 | 6 | 12 | 0.018475 | 0.221903 | 1.136357 | 0.012025 | 1.140042 | 1.008156 | 18.8278 | 12.83739 |
| Belgium | 36 | 33 | 11 | 23 | 0.004606 | 0.261187 | 1.164721 | 0.006302 | 1.352821 | 1.121602 | 545971.6 | 46.25027 |
| Brazil | 29 | 20 | 33 | 0 | 0.053807 | 0.065547 | 1.038249 | 0.004144 | 0.886084 | 0.782147 | 84868805 | 3162318 |
| Canada | 69 | 62 | 17 | 56 | 0.066305 | 0.15024 | 1.183832 | 0.013318 | 1.885461 | 1.397585 | 1028465 | 69300.79 |
| China | 20 | 18 | 8 | 17 | 0.099579 | 0.199368 | 1.119746 | 0.019578 | 1.091488 | 0.990578 | 121.6988 | 54.96871 |
| Denmark | 22 | 24 | 24 | 16 | 0.353108 | 0.848565 | 0.785263 | 0.180618 | 0.98842 | 1.445902 | 1.015964 | 1.012428 |
| France | 29 | 23 | 9 | 30 | 0.049381 | 0.110337 | 1.161091 | 0.004466 | 1.161717 | 1.015195 | 3700.744 | 272.5381 |
| Germany | 21 | 18 | 8 | 16 | 0.011451 | 0.068763 | 1.146057 | 0.00225 | 1.103203 | 0.978905 | 1681.327 | 728.2376 |
| India | 49 | 31 | 28 | 29 | 0.122032 | 0.137928 | 1.046921 | 0.014592 | 0.923097 | 0.84113 | 45051.62 | 15874.13 |
| Iran | 49 | 92 | 28 | 59 | 0.243367 | 0.2113 | 1.106169 | 0.062826 | 2.623699 | 1.902998 | 2.5E+16 | 1.07E+10 |
| Ireland | 34 | 34 | 15 | 25 | 0.50281 | 0.605517 | 1.076376 | 0.251619 | 0.98657 | 0.943088 | 288.961 | 133.1887 |
| Israel | 29 | 31 | 8 | 19 | 0.394938 | 0.583151 | 1.15689 | 0.070816 | 1.411518 | 1.254275 | 278.7508 | 164.3773 |
| Italy | 34 | 30 | 10 | 27 | 0.088719 | 0.184861 | 1.170778 | 0.004432 | 1.328466 | 1.121679 | 6024962 | 2.046924 |
| Malaysia | 25 | 33 | 11 | 27 | 0.869549 | 0.887035 | 1.128035 | 0.613072 | 3.326014 | 1.991332 | 462.0647 | 352.2444 |
| Netherlands | 43 | 48 | 12 | 26 | 0.209439 | 0.512291 | 1.161619 | 0.017553 | 1.493027 | 1.222879 | 1262766 | 2.636326 |
| Norway | 25 | 27 | 11 | 15 | 0.027648 | 0.701103 | 1.108444 | 0.050512 | 1.856421 | 1.301392 | 154.4844 | 124.6376 |
| Philipines | 51 | 53 | 28 | 36 | 0.182541 | 0.865493 | 1.01769 | 0.01217 | 3.117964 | 2.091286 | 68502.97 | 25666.16 |
| Poland | 48 | 77 | 23 | 68 | 1.022539 | 1.109518 | 1.132566 | 0.087518 | 2.135791 | 1.826874 | 1.02E+08 | 1.501466 |
| Portugal | 42 | 42 | 11 | 27 | 0.789265 | 0.698348 | 1.16253 | 0.743868 | 1.870177 | 1.538472 | 49310.62 | 1746.688 |
| Russia | 45 | 37 | 15 | 38 | 0.045172 | 0.045302 | 1.157039 | 0.004147 | 1.142701 | 0.997201 | 98785.97 | 3507.749 |
| South Korea | 14 | 14 | 5 | 7 | 0.688575 | 0.86664 | 1.127247 | 0.175176 | 1.656852 | 1.163943 | 363.2386 | 3.659551 |
| Spain | 24 | 19 | 8 | 21 | 0.071136 | 0.071543 | 1.166177 | 0.014662 | 1.172864 | 1.022933 | 2952.447 | 137.1358 |
| Sweden | 82 | 85 | 23 | 55 | 0.040168 | 1.051266 | 1.142619 | 0.006843 | 1.837028 | 1.534142 | 1.87E+10 | 634746.5 |
| Switzerland | 24 | 24 | 8 | 13 | 0.009232 | 0.446694 | 1.136035 | 0.021558 | 1.228293 | 1.083152 | 449.3075 | 32.22993 |
| Turkey | 27 | 26 | 9 | 24 | 0.320459 | 0.21737 | 1.135899 | 0.028505 | 1.153008 | 1.040987 | 20923327 | 139.9341 |
| UK | 44 | 39 | 14 | 34 | 0.054658 | 0.103414 | 1.14639 | 0.003097 | 1.327534 | 1.09207 | 196650.8 | 1463.103 |
| USA | 45 | 66 | 20 | 65 | 0.043499 | 0.027806 | 1.180173 | 0.502715 | 2.084341 | 1.49119 | 3.92E+08 | 3906578 |

[b] Good predictions are defined as the predicted result being within a factor of 2, predictions from 50 – 200% of the actual total number of cases). Thus, the quadratic model was able to predict the total number of cases in the United states in each of the 45 days before that day, while the exponential model was only within the defined good range for the 20 days preceding that day. The range of minimum predictions (under predictions) and maximum predictions (overpredictions) is also given

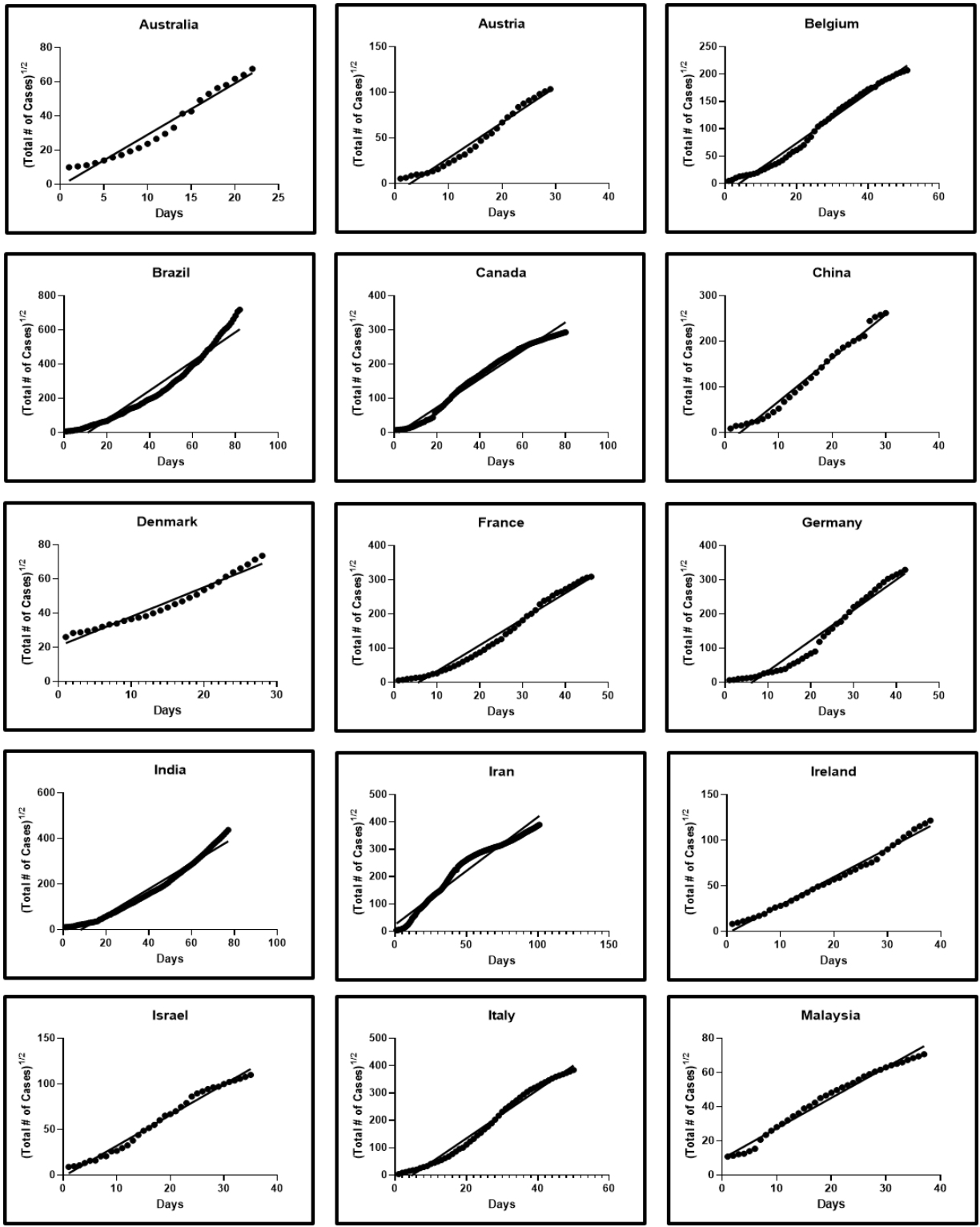

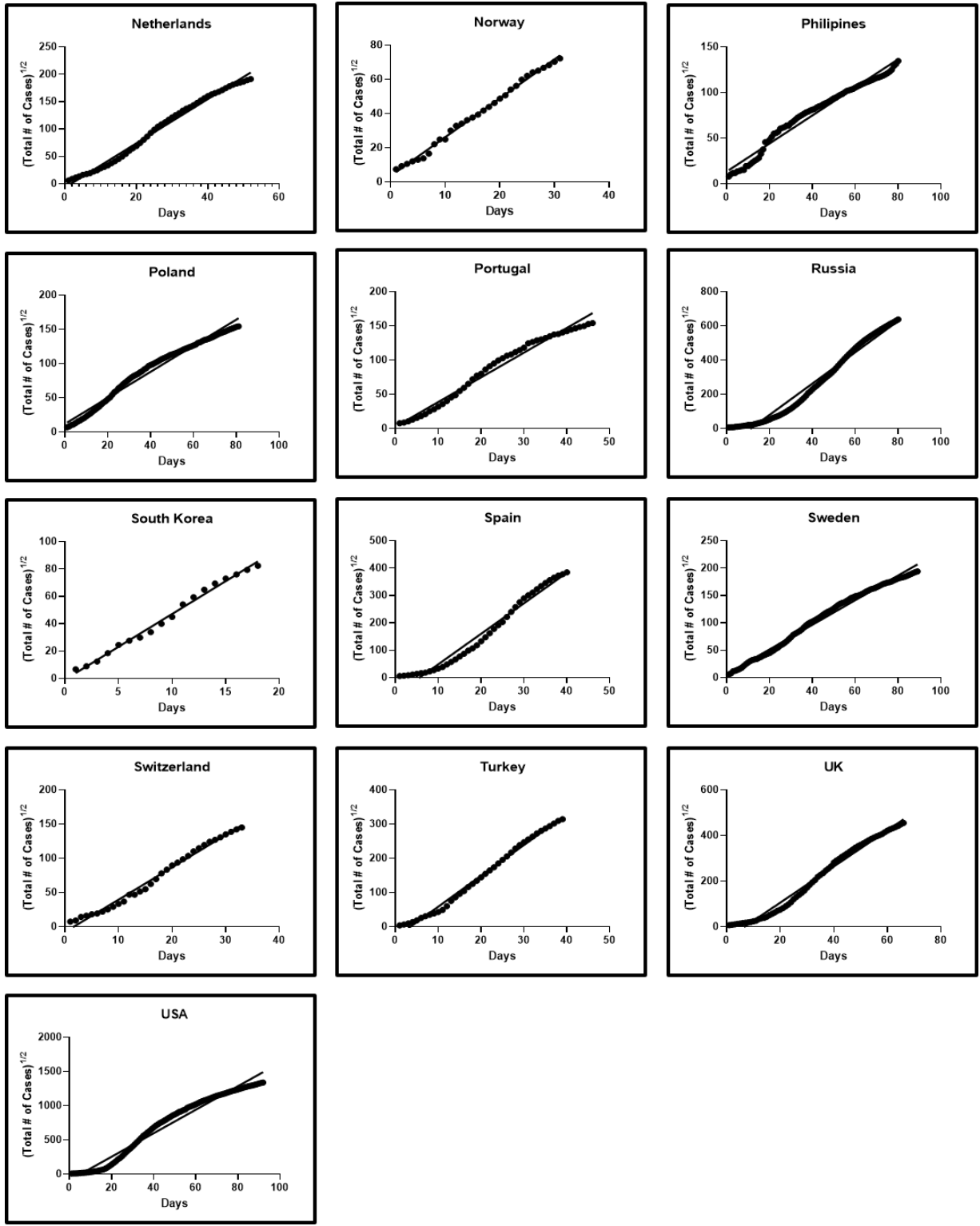

**SI Figure 1:** Plots of the square root of the total number of cases (√N) for the early portion of the COVID-19 epidemic in each of the 28 countries (black circles) along with a linear fit line.

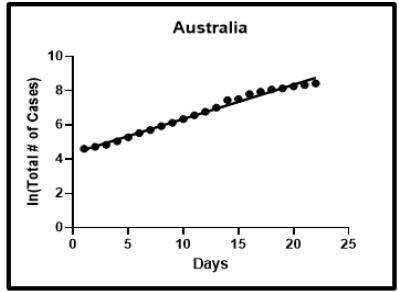
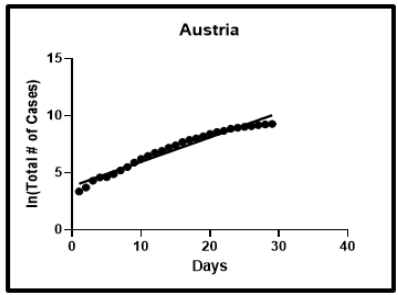
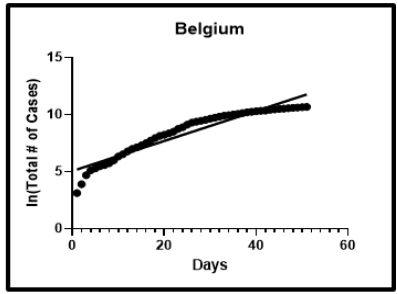
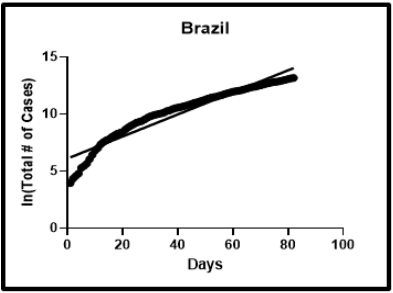
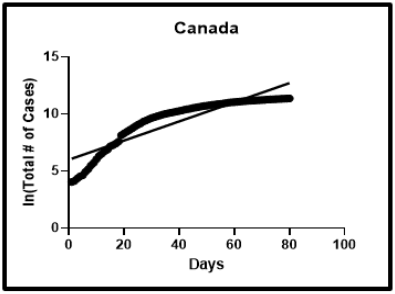
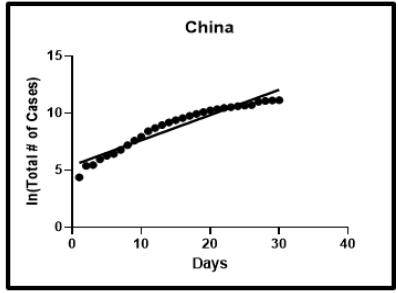
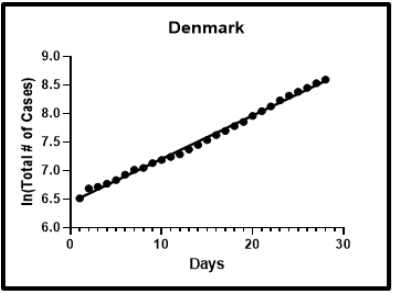
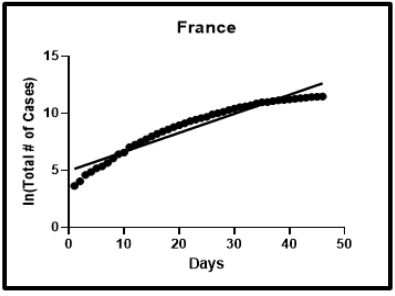
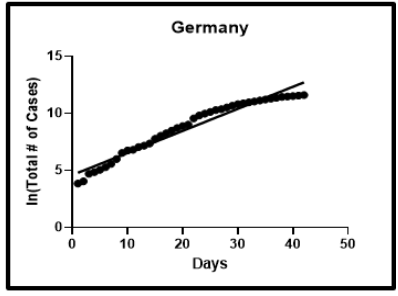
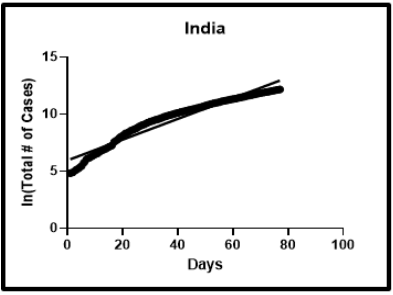
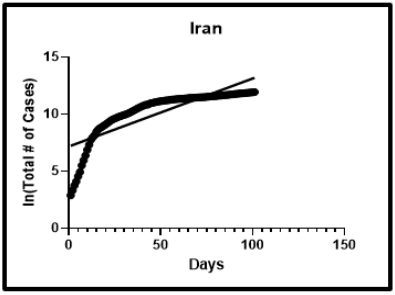
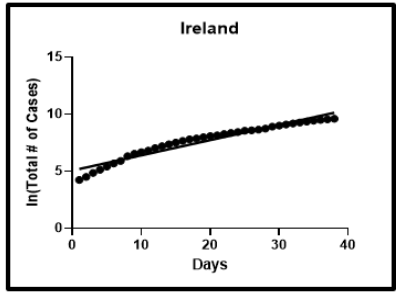
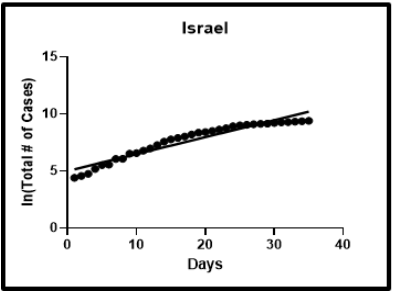
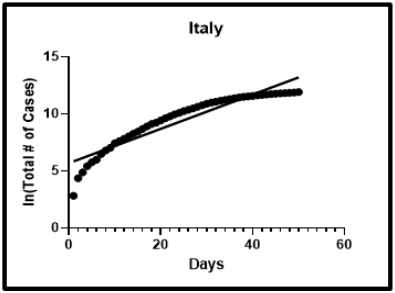
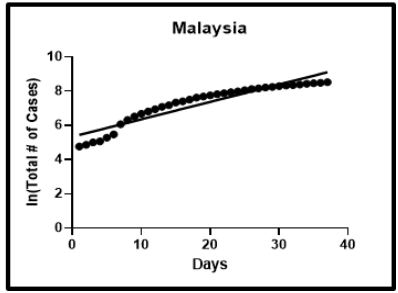

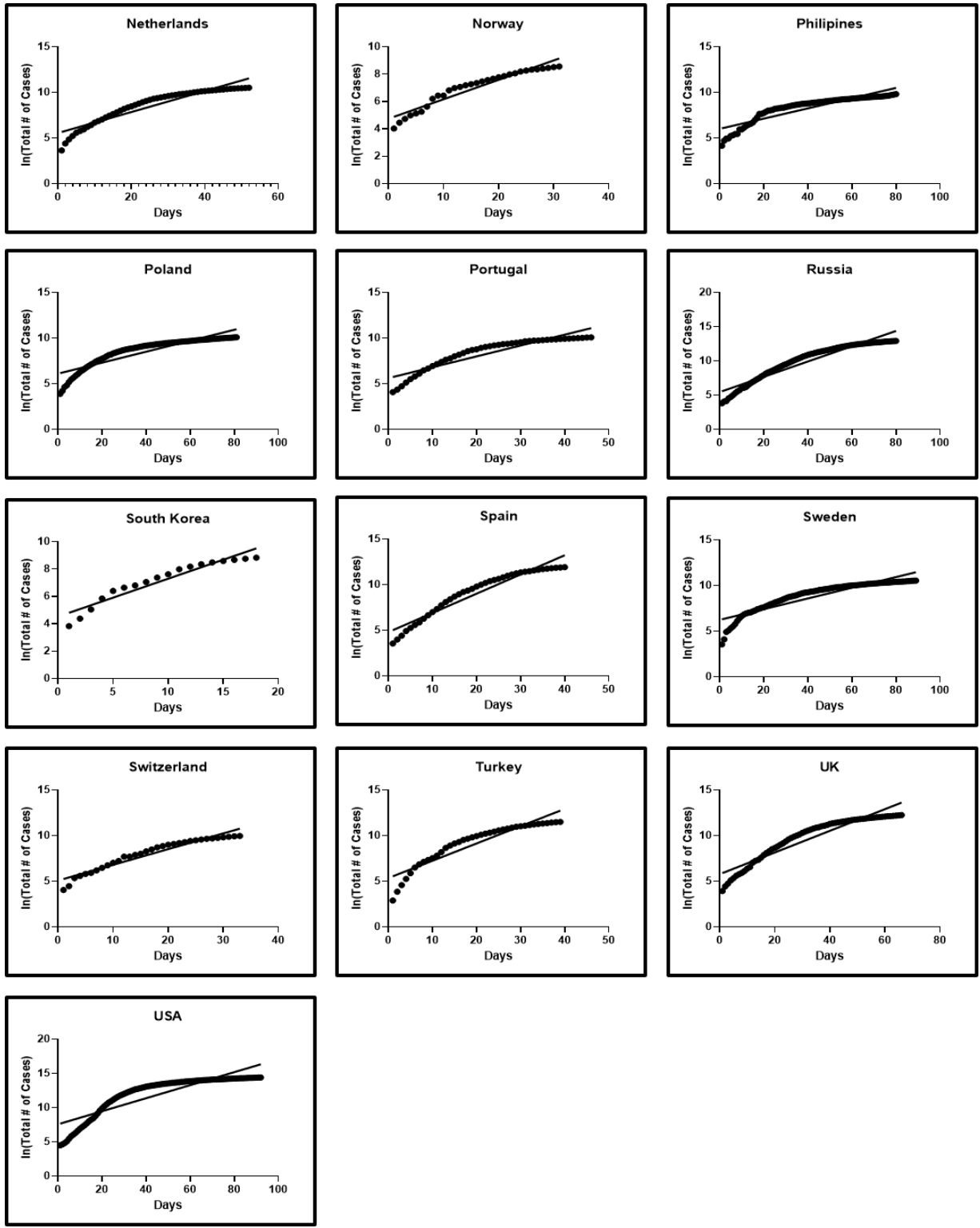

**SI Figure 2:** Plots of the natural log of the total number of cases (ln*N*) for the early portion of the COVID-19 epidemic in each of the 28 countries (black circles) along with a linear fit line.

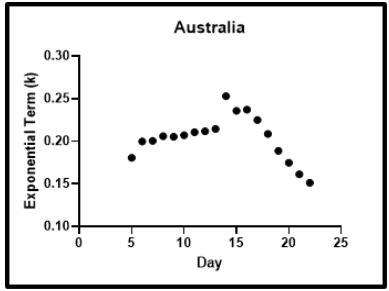
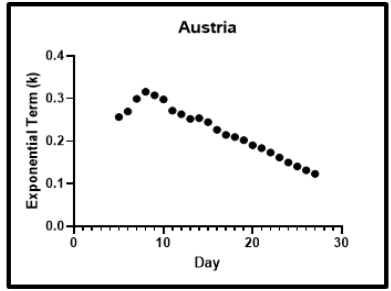
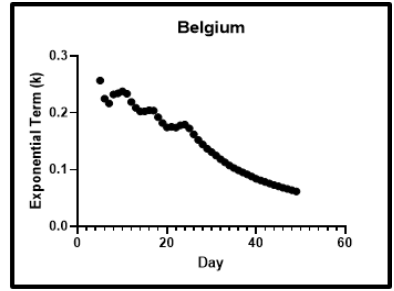
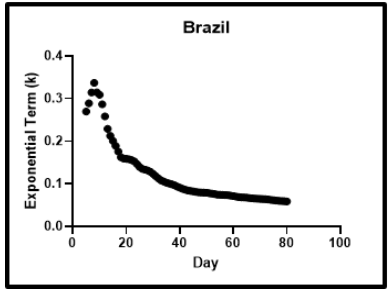
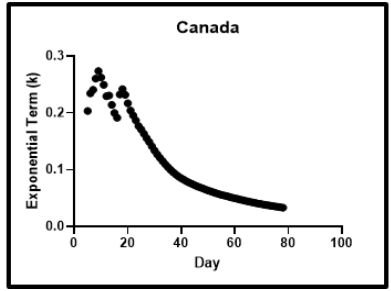
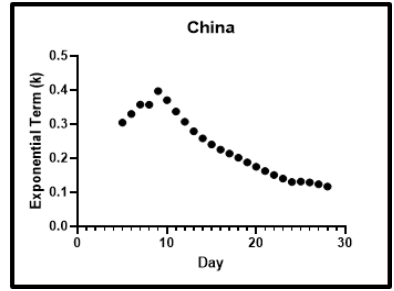
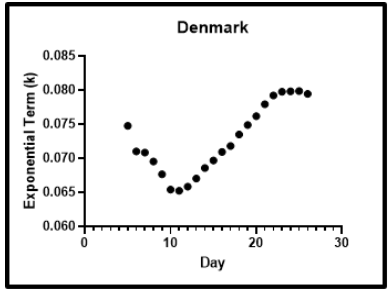
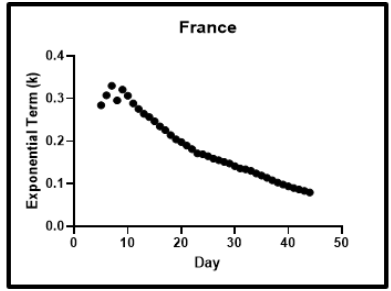
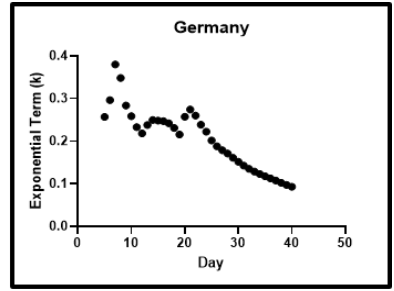
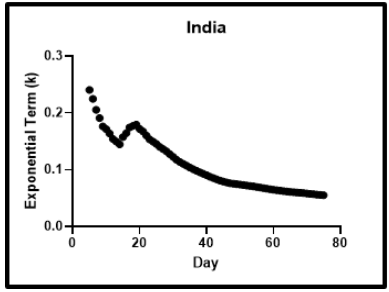
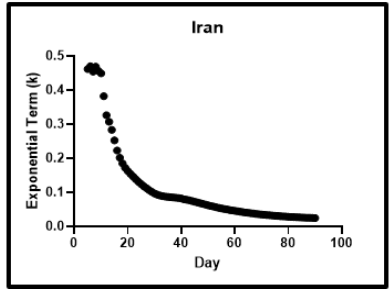
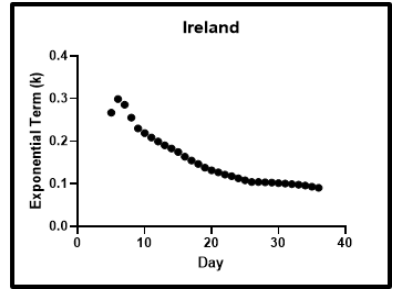
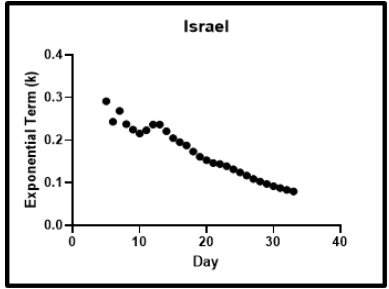
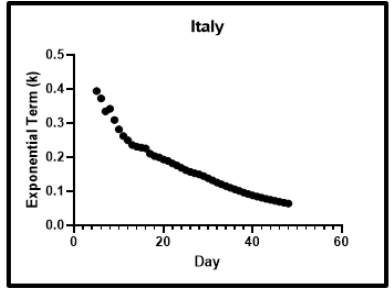
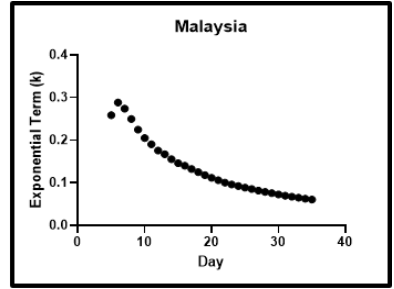
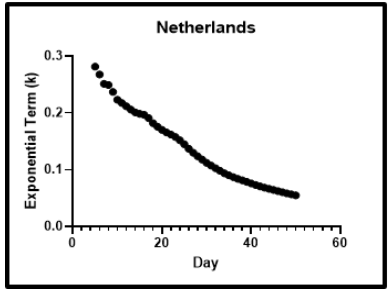
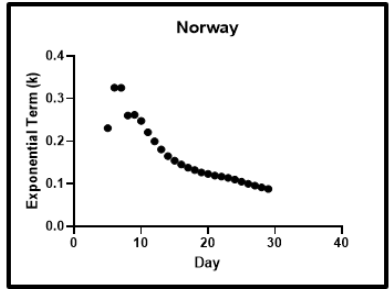
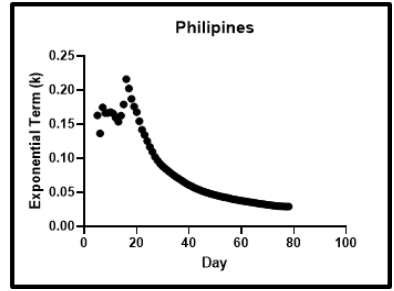

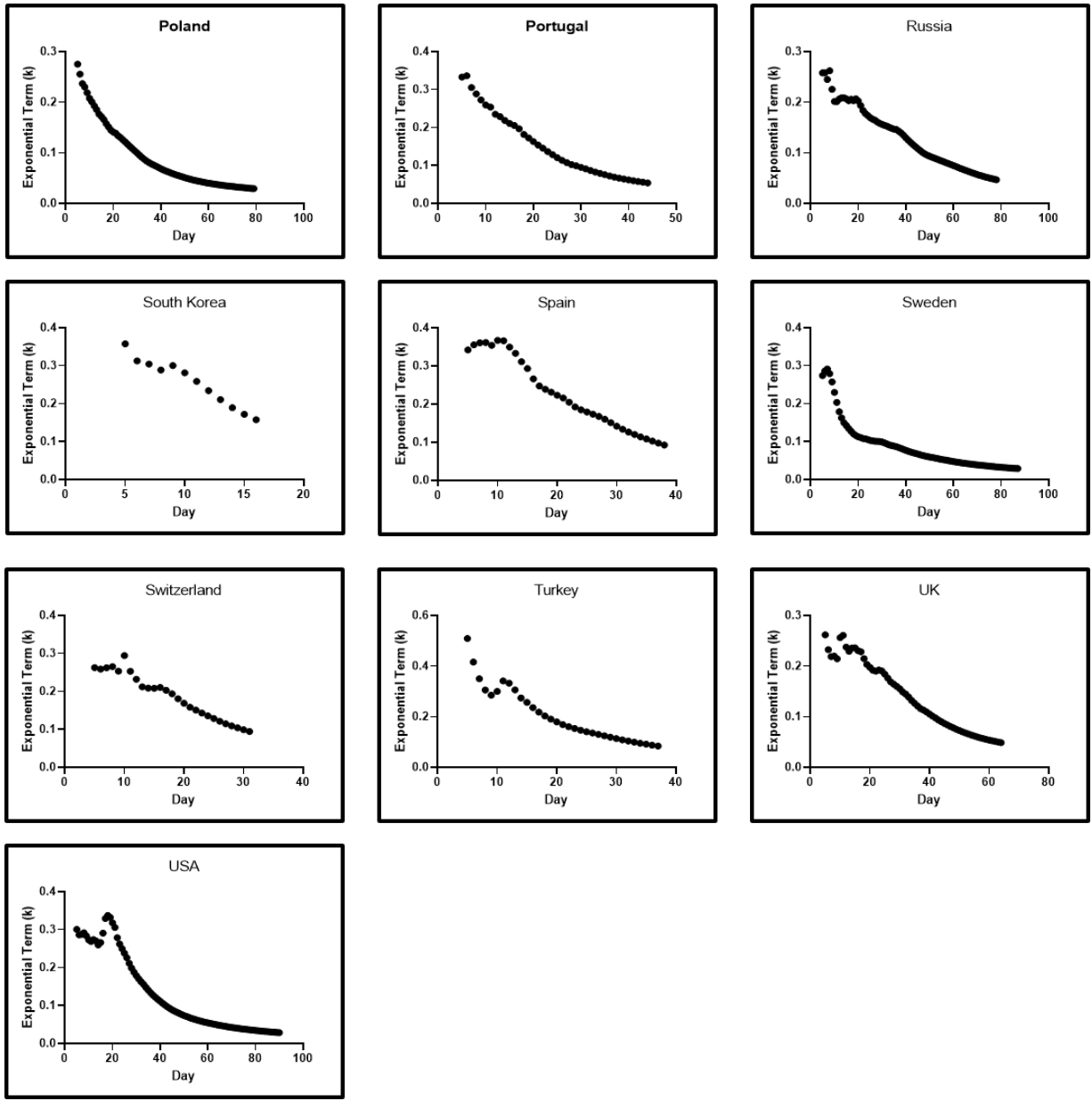

**SI Figure 3:** The change of the exponential rate term (k) over time for each of the 28 countries. It can be clearly seen that k is generally decreasing over time, often on each day but sometimes after an initial bit of increasing. This indicates that the exponential rate is regularly decreasing, as expected for a situation where growth resource is decreasing, as is expected for the logistic models family of models, including the generalized logistic and Gompertz growth models.